\documentclass[twocolumn]{aastex62}
\usepackage{float}
\usepackage{graphicx, graphics}
\usepackage{CJK}

\usepackage{amsmath, amssymb, bm}

\received{2017 December 22}
\revised{2018 March 16}
\accepted{2018 March 18}
\published{2018 April 12}
\submitjournal{The Astrophysical Journal Letters}

\shorttitle{Where are the MWC~758 arm drivers?}
\shortauthors{Ren et al.}

\begin{document}
\begin{CJK*}{UTF8}{gbsn}
\title{A Decade of MWC~758 Disk Images: Where Are the Spiral-Arm-Driving Planets?}
\author{Bin Ren (任彬)}\email{ren@jhu.edu} 
\affiliation{Department of Physics and Astronomy, The Johns Hopkins University, Baltimore, MD 21218, USA}

\author{Ruobing Dong (董若冰)}\email{rdong@email.arizona.edu}  
\altaffiliation{Bok Fellow}
\affiliation{Steward Observatory, University of Arizona, Tucson, AZ, 85719, USA}

\author{Thomas M.~Esposito}\email{tesposito@berkeley.edu}  
\affiliation{Department of Astronomy, University of California at Berkeley, Campbell Hall, Berkeley, CA 94720-3411}

\author{Laurent Pueyo}\email{pueyo@stsci.edu} 
\affiliation{Space Telescope Science Institute (STScI), Baltimore, MD 21218, USA} 

\author{John~H.~Debes} 
\author{Charles~A.~Poteet} 
\affiliation{Space Telescope Science Institute (STScI), Baltimore, MD 21218, USA}

\author{\'Elodie Choquet} 
\altaffiliation{Hubble Fellow}
\affiliation{Department of Astronomy, California Institute of Technology, 1200 E.~California Boulevard, Pasadena, CA 91125, USA}
\affiliation{Jet Propulsion Laboratory, California Institute of Technology, 4800 Oak Grove Drive, Pasadena, CA 91109, USA}

\author{Myriam Benisty} 
\affiliation{Unidad Mixta Internacional Franco-Chilena de Astronom\'{i}a, CNRS/INSU UMI 3386 and Departamento de Astronom\'{i}a, Universidad de Chile, Casilla 36-D, Santiago, Chile}
\affiliation{Universit\'e Grenoble Alpes, CNRS, IPAG, 38000 Grenoble, France}

\author{Eugene Chiang} 
\affiliation{Department of Astronomy, University of California at Berkeley, Campbell Hall, Berkeley, CA 94720-3411}
\affiliation{Department of Earth and Planetary Science, University of California at Berkeley, McCone Hall, Berkeley, CA 94720-3411}

\author{Carol A.~Grady} 
\affiliation{Exoplanets and Stellar Astrophysics Laboratory, Code 667, Goddard Space Flight Center Greenbelt, MD 20771, USA}

\author{Dean C.~Hines} 
\affiliation{Space Telescope Science Institute (STScI), Baltimore, MD 21218, USA}

\author{Glenn Schneider} 
\affiliation{Steward Observatory, The University of Arizona, Tucson, AZ 85721, USA}

\author{R\'emi Soummer} 
\affiliation{Space Telescope Science Institute (STScI), Baltimore, MD 21218, USA}

\begin{abstract}
Large-scale spiral arms have been revealed in scattered light images of a few protoplanetary disks. Theoretical models suggest that such arms may be driven by and co-rotate with giant planets, which has called for remarkable observational efforts to look for them. By examining the rotation of the spiral arms for the MWC 758 system over a 10-yr timescale, we are able to provide dynamical constraints on the locations of their perturbers. We present reprocessed {\it Hubble Space Telescope} ({\it HST})/NICMOS F110W observations of the target in 2005, and the new {\it Keck}/NIRC2 $L'$-band observations in 2017. MWC~758's two well-known spiral arms are revealed in the NICMOS archive at the earliest observational epoch. With additional {\it Very Large Telescope} ({\it VLT})/SPHERE data, our joint analysis leads to a pattern speed of $0\fdg6^{+3\fdg3}_{-0\fdg6}\, \mathrm{yr}^{-1}$ at $3\sigma$ for the two major spiral arms. If the two arms are induced by a perturber on a near-circular orbit, its best fit orbit is at $89$ au ($0\farcs59$), with a $3\sigma$ lower limit of 30 au ($0\farcs20$). This finding is consistent with the simulation prediction of the location of an arm-driving planet for the two major arms in the system.
\end{abstract}


\section{Introduction}
Planets form in gaseous and dusty protoplanetary disks around young stars a few million years old. Forming planets gravitationally interact with the host disk, producing structures such as gaps, spiral arms, and vortices \citep{kley12}. By comparing observations with theoretical models, spatially resolved disk structures may yield rich information about the properties of embedded planets, such as their orbits, and dynamical constraints on their masses. 

In the past decade, near-infrared imaging of disks with high spatial resolution has discovered spiral arms at tens of au in a few systems (e.g., SAO~206462: \citealp{muto12, garufi13, stolker16}; LkH$\alpha$~330: \citealp{akiyama16}; MWC~758: \citealp{grady13, benisty15};  HD~100453: \citealp{wagner15, benisty17}; {and HD~141569~A: \citealp{mouillet01, clampin03, konishi16}}). Hydrodynamical and radiative transfer simulations have suggested two mechanisms for reproducing such structure: gravitational instability \citep{lodatorice15, dong15b} which occurs in disks with sufficient mass \citep{kratter16}, and companion-disk interaction \citep{dong15, zhu15, bae16}. Because the host disks in these few systems are probably not massive enough to trigger the gravitational instability \citep[e.g., ][]{andrews11}, the latter scenario is more likely. 

Detailed numerical simulations have quantified the dependencies of arm separation and contrast on the companion mass and disk properties \citep{fung15, dong17}. These relations have been used to infer the parameters of hypothesized arm-driving companions. A proof of concept of this mechanism is recently provided by the HD~100453 system, where both the arms and the companion have been found, with their physical connections numerically supported \citep{dong16, wagner18}. Extensive direct imaging observations have been carried out to look for the predicted arm-driving companions in a few other systems. Assuming hot start planet formation models \citep[e.g.,][]{baraffee15}, they have generally ruled out all but planetary mass objects of a few Jupiter masses or less \citep[e.g., ][]{maire17}.

Companion-driven arms co-rotate with their driver. Therefore, by measuring their pattern speed, the orbital period, thus semi-major axis, of their companion can be constrained \citep[e.g.,][]{lomax16}. We perform such an exercise for the spiral arm system MWC~758, taking advantage of observations of the arms over a decade-long baseline established by a 2005 {\it HST}/NICMOS observation and 2015/2017 {\it VLT}/SPHERE and {\it Keck}/NIRC2 observations. 

MWC~758 is a Herbig Ae star located at $151_{-9}^{+8}$ pc \citep{gaia16} with an age of $3.5\pm2.0$ Myr \citep{meeus12}, and mass of  $\sim2.0 M_{\odot}${\footnote{We derive the stellar mass from the \citet{siess00} pre-main sequence evolutionary tracks, assuming stellar effective temperature and luminosity of 7580K and 19.6 $L_\odot$ (\citealp{vanderMarel16}; after scaling the stellar luminosity using the new {\it Gaia} distance).}} . The disk has a low inclination of ${\sim}20^\circ$ \citep{isella10}. Its two prominent, roughly $180^\circ$ rotationally symmetric arms were first discovered with {\it Subaru}/HiCIAO \citep{grady13}, and subsequently characterized in detail with {\it VLT}/SPHERE \citep{benisty15}, with a third arm and point-source candidate {at ${\sim}0\farcs11$ (17 au)} recently reported in \citet{reggiani18} using {\it Keck}/NIRC2. Numerical simulations by \citet{dong15} suggested that both arms can be produced by a multi-Jupiter-mass planet at $\sim0\farcs6$ from the star.

\section{Data Acquisition \& Reduction}
In this Section, we describe the observations and data reduction for our 2005 {\it HST} and 2017 {\it Keck} programs.

\subsection{NICMOS}
The {\it HST}/NICMOS coronagraphic instrument observed the MWC~758 system in total intensity with the F110W filter ($\lambda_{\text{cen}}=1.1\ \micron$) on 2005 January 7 (Proposal ID: 10177, PI: G.~Schneider), and the unresolved disk morphology was presented in \citet{grady13}. To retrieve the morphology of the spiral arms, we obtain calibrated NICMOS images of MWC~758 and another 814 reference star exposures, i.e., point-spread functions (PSFs), from the Archival Legacy Investigations for Circumstellar Environments (ALICE) project (PI: R.~Soummer; \citealp{choquet14,hagan18}). We align the observations for better astrometry by employing a Radon-transform-based technique (\citealp{pueyo15}; C.~Poteet et al., ApJ submitted), which focuses on the diffraction spikes in each exposure. To minimize color mismatch, telescope breathing and cold mask alignment, we select 81 closest PSFs in the $L^2$-norm sense, and perform PSF subtraction with the Non-negative Matrix Factorization (NMF) method \citep{ren18}, which is shown to preserve the morphology of circumstellar disks better, especially in reference differential imaging scenarios.

\begin{figure}[htb!]
\center
\includegraphics[width=.49\textwidth]{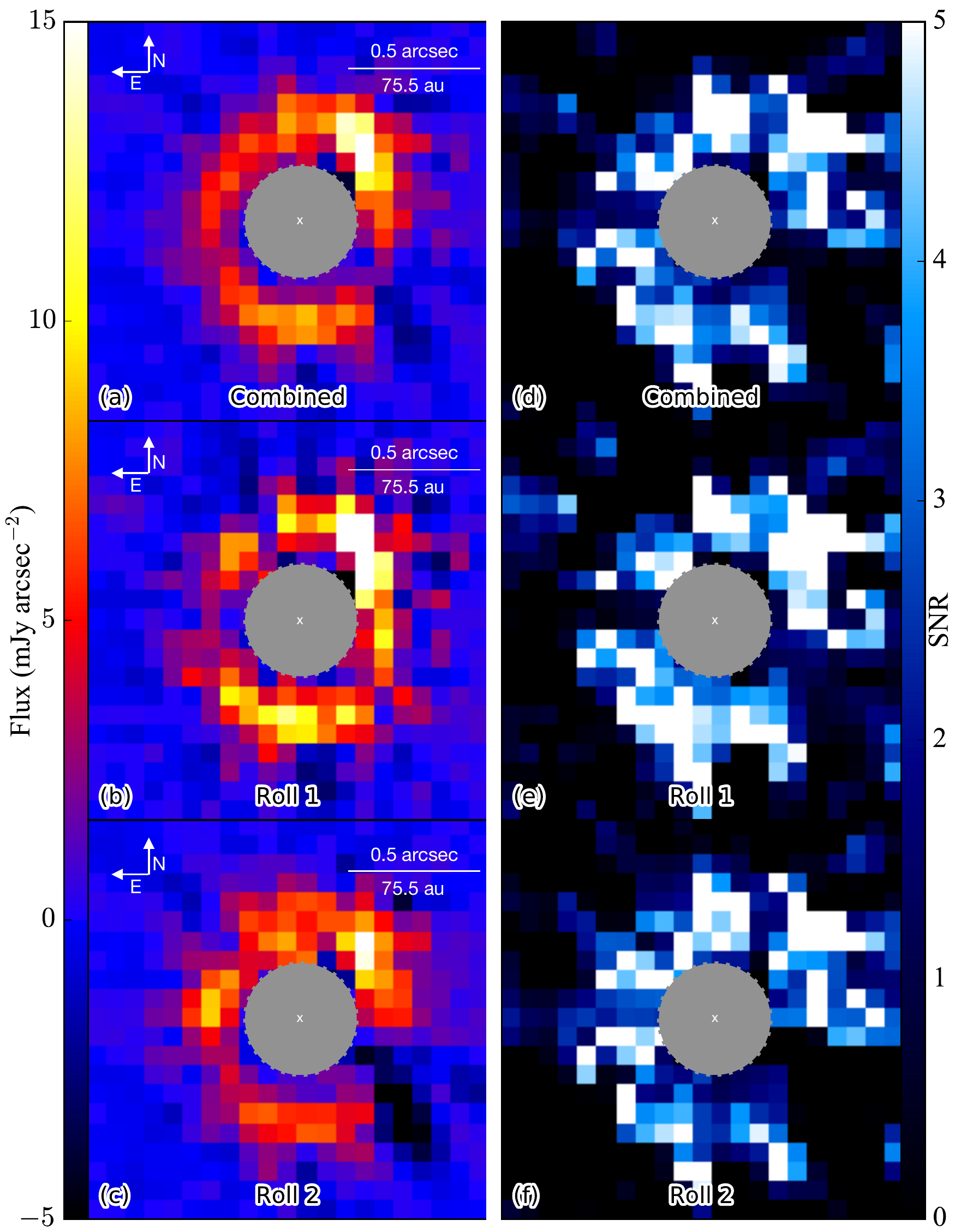}
\caption{{\it Left}: The NICMOS images of MWC~758, including the combined ({\bf a}) and two different rolls in ({\bf b}, {\bf c}). {\it Right}: The S/N maps, calculated from dividing the final images by the pixel-wise standard deviation of their constituting ones. The inner working angles are marked with gray circles, and stellar locations with white crosses.}
\label{fig:raw}
\end{figure}

In Fig.~\ref{fig:raw}, we present the reduction results of the NICMOS observations at two telescope orientations (three exposures each) and their signal-to-noise (S/N) maps. We argue the physical existence of the detection since the spiral pattern is (1) consistent within the same telescope orientation, as shown from the S/N maps which are calculated from dividing the combined result by the pixel-wise standard deviation of the ones constituing them; (2) independent of telescope orientation ($30^\circ$ separation), which excludes the scenario of unsuccessful noise removal; (3) not depending on the number of NMF components, reducing the possibility of underfitting and overfitting; (4) not resembling the null detections in the ALICE archive, as well as a reduction consistency using a principal-component-analysis-based reduction method \citep{soummer12}.

The integrated flux for $0\farcs3<r<0\farcs5$ is $2.0\pm0.8$ mJy at $1\sigma$ level, consistent with the upper limit reported in \citet{grady13}. We notice flux variations between the two telescope orientations, however we do not address the origin of this difference in this letter, but focus on the morphology only.

\begin{figure*}[htb!]
\center
\includegraphics[width=.95\textwidth]{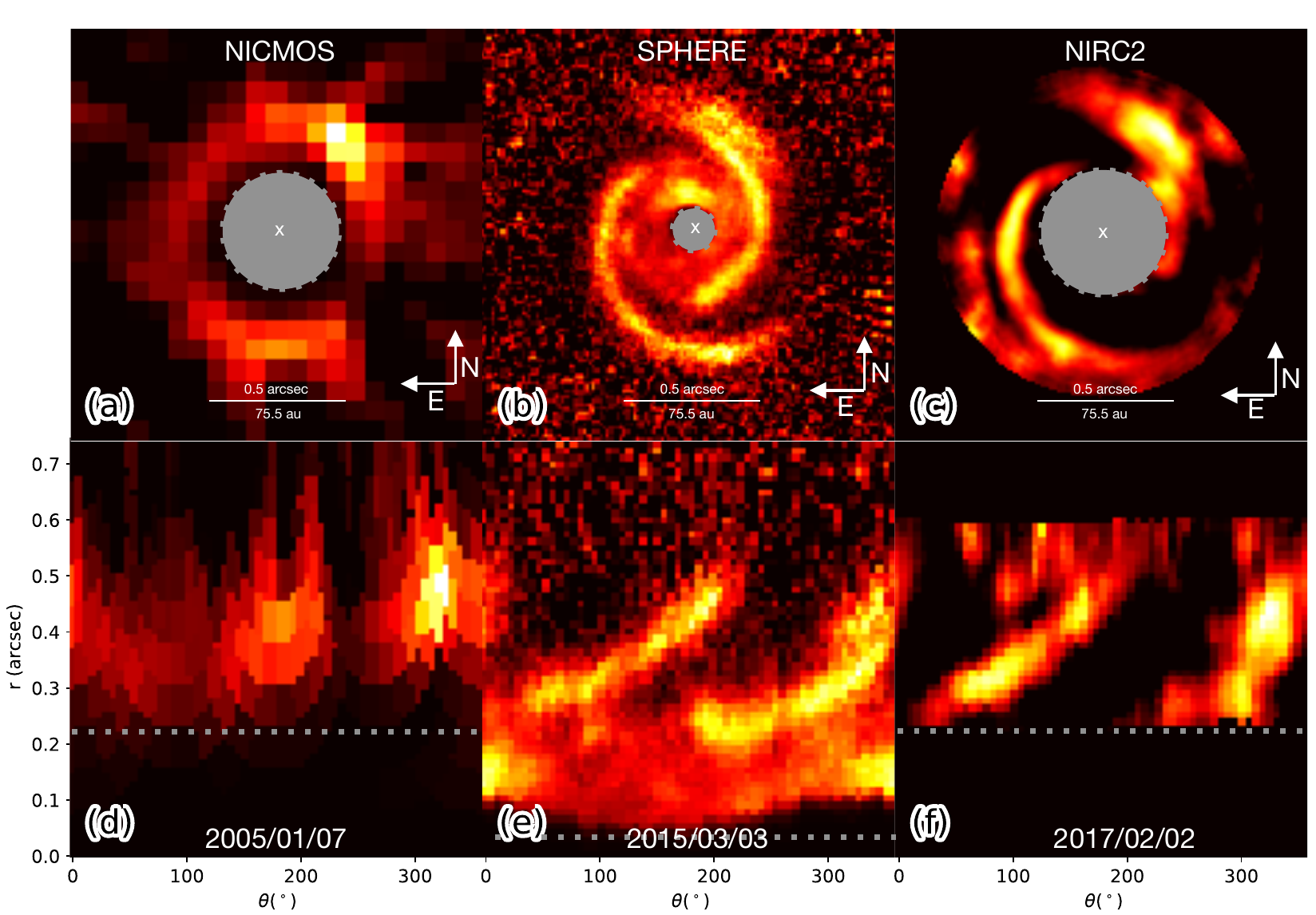}
\caption{The $r^2$-scaled NICMOS, SPHERE, and NIRC2 (from left to right) observations of MWC~758 in Cartesian (top) and polar (bottom) coordinates with total flux normalized to unity. The gray dotted circles and lines mark the inner working angles.}
\label{fig:carpol}
\end{figure*}

\subsection{NIRC2}\label{nirc2reduction}
We observed MWC~758 with {\it Keck}/NIRC2 in $L'$-band total intensity ($\lambda_{\text{cen}}=3.8\ \micron$) on 2017 February 2 (PI: E.~Chiang). The data were obtained with the narrow camera (9.971 mas pixel$^{-1}$; \citealp{service16}) in ``vertical angle mode'' to allow for angular differential imaging (ADI; \citealp{marois06}). Our observations totaled 262 images, each consisting of 30 coadds of 1.0-s exposures, covering 161$^\circ$ of field rotation. Airmass varied from 1.01 to 1.39 and precipitable water vapor was approximately 2.5 mm.


The vector vortex coronagraph \citep{serabyn17} was used in combination with the existing Keck II adaptive optics system to suppress host star light. The QACITS control system \citep{huby17} maintained alignment of the vortex mask with the star during observations, and images are aligned with each other to sub-pixel precision using a downhill simplex algorithm to minimize residuals of the stellar PSF in frames differenced with a selected reference frame. Calibrated images are produced from raw images by performing dark subtraction, flat-fielding, thermal background subtraction with dedicated sky frames, and distortion correction \citep{service16}. The absolute star positions are then determined to 0.5 pixel precision in both spatial dimensions by a Radon transform of the averaged frames \citep{pueyo15}.

We subtract the stellar PSF from the calibrated images using the NMF method. PSF-subtraction algorithms with ADI are known to distort the morphology of extended objects from self-subtraction \citep[e.g.,][]{follette17}; therefore, for each image, this bias is avoided by a minimum rotation threshold of $45^\circ$ for the selection of its reference images; the final result is then the median of the PSF-subtracted images.

\subsection{SPHERE}
We obtain the {\it VLT}/SPHERE polarized intensity result in $Y$-band ($\lambda_{\text{cen}}=1.04\ \micron$) on 2015 March 3 from \citet{benisty15}.

\section{Data Analysis}\label{analysis}
To measure the pattern speed of the arms, we first scale the surface brightnesses of the reduced NICMOS, NIRC2, and SPHERE images by the distance-dependent factor $r^2$. The results in Cartesian and polar coordinates are presented in Fig.~\ref{fig:carpol}.

There are three main differences among our observational datasets: (1) the pixel size of the NICMOS instrument is ${\sim}8$ times larger than the other two; (2) the NICMOS and NIRC2 observations measure the total intensity while SPHERE traces the polarized light, and (3) the NICMOS and SPHERE observations are at ${\sim}1\ \micron$ while the NIRC2 observation is at ${\sim}3.8\ \micron$. For (1), we interpolate the NICMOS image to match the pixel scales of the others. For (2), we reduced the SPHERE ADI total intensity observation, compared it with the polarized map, and found no discernible discrepancy; this agreement is also endorsed by simulation in \citet{dong16b}. For (3), we compare observations at roughly the same central wavelength ($\lambda_{\text{cen}}$).

\subsection{Measurement of Rotation of Spirals}\label{rotation_measured}
\subsubsection{NIRC2: 2015 vs 2017 (1.28 yr apart)}\label{consistency}
To mitigate any systematic offset between instruments, and provide an initial constraint on the rotation, we compare two observations from the same {\it Keck}/NIRC2 instrument: our 2017 February 02 observation, and the 2015 October 24 observation (PI: E.~Serabyn, \citealp{reggiani18}) which is aligned and reduced with identical procedure.

\begin{figure*}[htb!]
\center
\includegraphics[width=\textwidth]{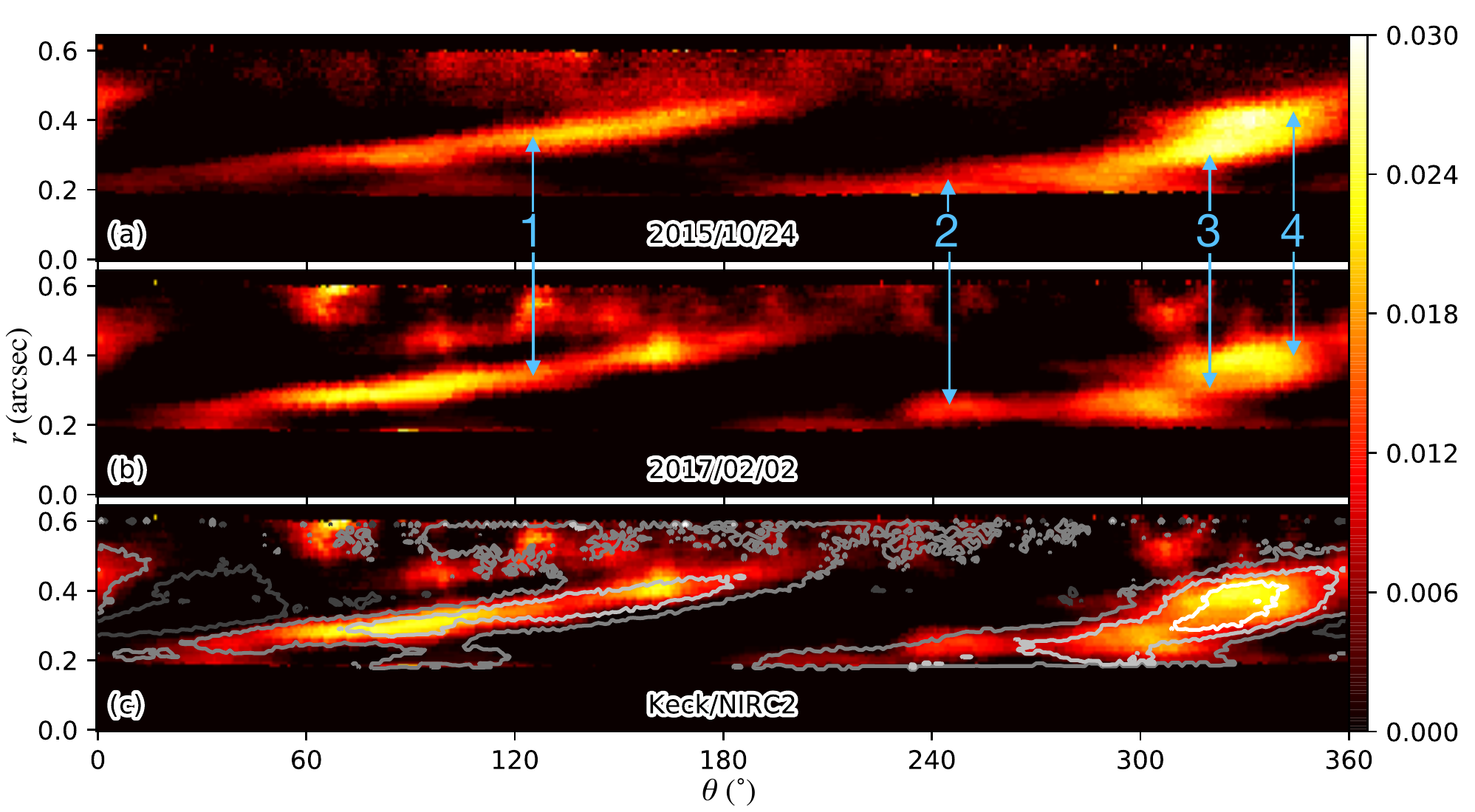}
\caption{MWC~758 spiral arms observed by {\it Keck}/NIRC2 in polar coordinates (total flux normalized to 1 for comparison) in 2015 ({\bf a}) and 2017 ({\bf b}), and contour of 2015 observation overplotted on the 2017 one ({\bf c}).}
\label{fig:nirc2}
\end{figure*}

We quantify the rotation of the spiral arms as follows: in polar coordinates, we first fit Gaussian profiles to the brightness of the spiral arms at a fixed radial separation; then for each arm, we perform the weighted Least Square Dummy Variable (LSDV, Appendix~\ref{lsdv}) analysis to fit the same morphological profile in both epochs to obtain their relative rotation. For the Southern and Northern primary arms (marked with ``1'' and ``3'' in Fig.~\ref{fig:nirc2}), we obtain a rotation of $\Delta\theta_S^{(1.28\,\mathrm{yr})}=0\fdg77\pm10\fdg65$\footnote{The calculated uncertainty in this letter is $3\sigma$ unless otherwise specified.}, and $\Delta\theta_N^{(1.28\,\mathrm{yr})}=-0\fdg70\pm6\fdg78$, respectively. Since spiral arms in disks are trailing patterns, the MWC~758 arms are expected to rotate in a clockwise direction, i.e., $\Delta\theta\geq0$, we therefore adopt the constraints from the Southern primary arm, $\dot\theta=\Delta\theta_S^{(1.28\,\mathrm{yr})}\big/1.28\mathrm{yr}=0\fdg6\pm8\fdg3$ yr$^{-1}$, as the rotation of the two.

For consistency check, we measure the rotation with another method: in polar coordinates, we obtain the cross-correlate maps \citep{tonry79},  and measure $\dot\theta=0\fdg7\pm56\fdg2$ yr$^{-1}$. We adopt the value from the LSDV method, since it is less biased by the non-spiral structures in the entire field of view as the cross-correlation method, and the best-fit values agree within $0\fdg1$.

\subsubsection{2005 NICMOS vs 2015 SPHERE (10.17 yr apart)}\label{quantify}
By analyzing the rotation between the NICMOS and SPHERE images, we narrow down the uncertainty determined from the NIRC2 observations here. We adopt the NIRC2 best-fit and uncertainty values, rotate the SPHERE result back to the NICMOS epoch, then the two images are expected to have no azimuthal shift. By fitting identical profiles and offsets for the rotated SPHERE and original NICMOS observations, and given the Northern arm is blended with its secondary arm (marked by ``4'' in Fig.~\ref{fig:nirc2}) but with a smaller uncertainty than the Southern arm, we adopt the results from the Southern arm, obtaining a conservative measurement of $\Delta\theta^{(10.17\,\mathrm{yr})}=6\fdg1\pm29\fdg4$. This corresponds to a  statistical uncertainty for the angular speed: $(\delta\dot\theta)_{\mathrm{statistical}} = 29\fdg4\big/10.17\mathrm{yr}= 2\fdg89$ yr$^{-1}$.

\subsection{Additional Systematics}\label{rotation_systematics}
We identify and study the impact of two possible systematics associated with the NICMOS results: the alignment uncertainty of stellar center determination (CD), and the misalignment uncertainty of the star behind the focal plane mask (FPM).

To quantify the stellar center determination uncertainty, we cross-correlate the raw MWC~758 exposures with the $814$ ALICE references, and determine the $3\sigma$ uncertainty to be $0.5$ pixel along both horizontal and vertical directions. We then draw $1,000$ possible centers within $\pm 0.5$ pixel from the center determined by our Radon Transform method, and cross-correlate the arm images in polar coordinates with the SPHERE result, and obtain a $3\sigma$ quantile of $\Delta\theta_{\mathrm{CD}}=0.6^\circ$. We therefore adopt a $3\sigma$ upper limit of {$\delta\dot\theta_{\text{CD}}=0.6^\circ\big/10.17\mathrm{yr}=0\fdg06$ yr$^{-1}$}.

Since the arms lie near the edge of the NICMOS FPM, if they do have rotated in this $\sim$10-yr span, with the star not well-centered on the FPM during the time of the observation, this may still yield nonsignificant moving spirals. To account for this, we simulate $1,000$ SPHERE images with the following two parameters: (1) rotations within $\pm60^\circ$ (denoted as $\alpha$): a range that the arms would rotate in ${\sim}10$ years if they are driven by the protoplanet candidate reported by \citet{reggiani18}, and (2) shifted centers within $\pm 0\farcs04$ ($0.5$ NICMOS pixel) along both horizontal and vertical directions. We mask the resampled SPHERE data with a circle of the NICMOS FPM size, then cross-correlate them with the original SPHERE image, and obtain their relative azimuthal shift ($\Delta\theta'$), which is then subtracted by introduced shift ($\alpha$). We obtain a $3\sigma$ upper limit $\delta(\Delta\theta)_{\text{FPM}} = \Delta\theta' - \alpha = 11^\circ$. {Therefore, $(\delta\dot\theta)_{\mathrm{FPM}}=11^\circ\big/10.17\mathrm{yr}=1\fdg08$ yr$^{-1}$.}\\

\section{Result}\label{result}
From the previous analyses of statistical and systematical uncertainties, we obtain a total uncertainty in the rotation of the arms at $3\sigma$:
\begin{align*}
&\delta\dot\theta= \sqrt{(\delta\dot\theta)^2_{\text{statistical}}+(\delta\dot\theta)^2_{\text{systematic}}}\\
&= \sqrt{(\delta\dot\theta)^2_{\text{statistical}}+\left[(\delta\dot\theta)^2_{\text{CD}}+(\delta\dot\theta)^2_{\text{FPM}}+(\delta\dot\theta)^2_{\text{pixel}}\right]}\\
&= \sqrt{2.89^2+(0.06^2+1.08^2+1.18^2)} = 3\fdg31\, \mathrm{yr}^{-1},
\end{align*}where the pixel uncertainty is accounted for the NICMOS pixel size of $12^\circ$ at $r{\sim}0\farcs6$ (i.e., $\delta\dot\theta_{\text{pixel}}=12^\circ\big/10.17\mathrm{yr}=1\fdg18$ yr$^{-1}$). Together with the best-fit value, we obtain \begin{equation}
\dot\theta=0\fdg6^{+3\fdg3}_{-0\fdg6}\, \mathrm{yr}^{-1},
\end{equation}
where the lower limit is physically constrained from the clock-wise rotation of the MWC~758 arms.

\begin{figure}[htb!]
\center
\includegraphics[width=.45\textwidth]{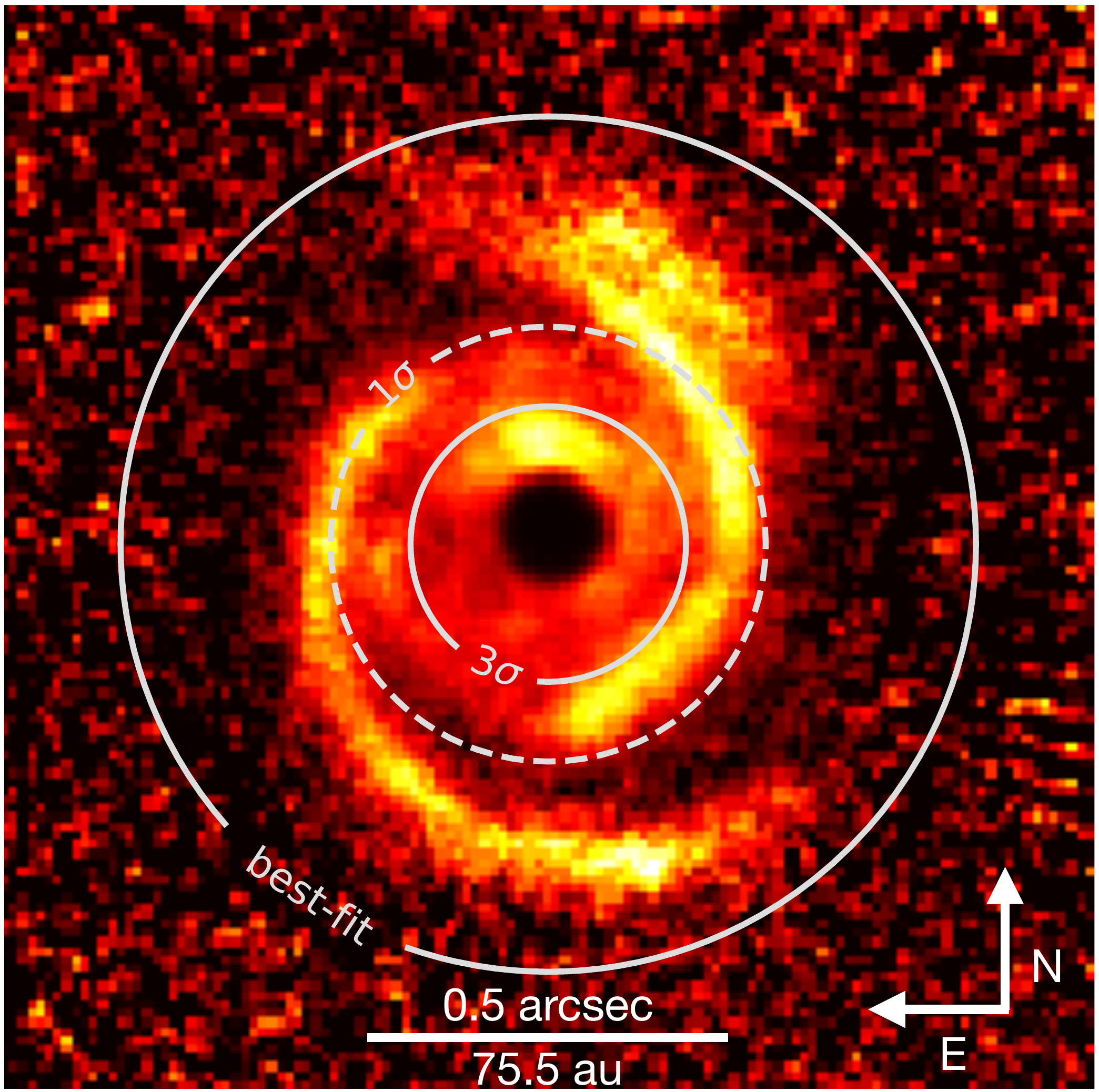}
\caption{{The best-fit and $3\sigma$ lower limit (solid lines) of where the major-arm-driving planets should be based on the measured pattern speed, plotted over the SPHERE data. The $1\sigma$ limit (dashed line), which is interpolated from the best-fit and the $3\sigma$ limit, is presented for illustration purpose and it may not represent a true 68\% likelihood.}}
\label{fig:planet}
\end{figure}

For hypothesized arm-driving planet(s) on a circular orbit (eccentricity $e=0$), the best-fit pattern speed corresponds to a period of $T=598$ yr, or a radial separation of $r_{\text{best}} = 89$ au; and the $3\sigma$ upper limit leads to $T=92$ yr and $r_{3\sigma}=26$ au; see Fig.~\ref{fig:planet} for graphical representations.

For $e>0$, if the planet reaches its apogee in the current epoch, the limit on the arm motion translates into a stellocentric separation $r_{\text{min}}(e) = (1-e)^{1/3}r$ au. For giant planets with several Jupiter mass forming in and interacting with a gaseous disk, their eccentricities are not expected to grow beyond $\sim$0.1 \citep{dunhill13, duffell15}. Furthermore, density waves excited by Jovian planets with $e\gtrsim0.2$ start to deviate from their usual morphology, as the waves launched at different epochs interact with each other (Hui Li \& Shengtai Li, private commnication), which provide poor fits to the arms around MWC~758. In this case, $e=0.2$ leads to a $7\%$ decrease in the minimum stellocentric separation, and the hypothetical arm-driving planet(s) should be located at least $25$ au from the star in 2017 at $3\sigma$.

One might consider another scenario in which the two spiral arms exchanged positions between 2005 and 2015, i.e., rotated ${\sim}180^\circ$. However, this means the major arms should have rotated an additional $22\fdg6$ between the 2015 and 2017 NIRC2 observations, this is ruled out at $6\sigma$ with our constraints.

\section{Summary and Discussion}

We present reprocessed 2005 {\it HST}/NICMOS observations of the MWC 758 disk, and successfully retrieve the two spiral arms in the system revealed by ground-based high-contrast imaging facilities. Thanks to the {\it HST} image, we are able to establish a 10-yr baseline in observations to constrain the pattern speed of the major arms. Together with a 2015 {\it VLT}/SPHERE dataset, and two {\it Keck}/NIRC2 observations in 2015 and 2017, we obtain a rotation speed of {$0\fdg6^{+3\fdg3}_{-0\fdg6}\, \mathrm{yr}^{-1}$ at $3\sigma$ for the two major spiral arms. The results correspond to a best-fit value of $89$ au ($0\farcs59)$, and a $3\sigma$ lower limit of $26$ au ($0\farcs17$), for the orbital distance of the hypothesized arm-driving perturber on a circular orbit.}

Our constraint is consistent with the  \citet{dong15} prediction of the arm-driving planet at ${\sim}90$ au (${\sim}0\farcs6$). In addition, we rule out at a $6\sigma$ level the scenario that the companion candidate at $0\farcs11$ (17 au) reported by \citet{reggiani18} is driving the major two spiral arms, assuming the candidate is on a circular orbit coplanar with the arms. This is further supported by \citet{bae18}, that a Jovian planet can drive only one external arm assuming a reasonable disk scale height. For new arm(s) and planet candidate(s) \citep[e.g.,][]{reggiani18}, more observations are needed to confirm their existence and dynamical connections.

The possible arm-driving planets in the MWC~758 system are excellent targets for future observations in direct imaging observations both from the ground and with the {\it James Webb Space Telescope}, and in millimeter observations using {\it ALMA} to search for evidence of a circumplanetary disk \citep[e.g.,][]{zhu15a, eisner15, perez15, szulagyi17}.

\acknowledgments
We are grateful to an anonymous referee for constructive suggestions that improved our letter, and Sean Brittain for insightful discussions. B.R.~acknowledges the computational resources from the Maryland Advanced Research Computing Center (MARCC), which is funded by a State of Maryland grant to Johns Hopkins University through the Institute for Data Intensive Engineering and Science (IDIES). E.C.~acknowledges support from NASA through Hubble Fellowship grant HST-HF2-51355 awarded by STScI, operated by AURA, Inc.~under contract NAS5-26555, and support from HST-AR-12652, for research carried out at the Jet Propulsion Laboratory, California Institute of Technology. T.E.~was supported in part by NASA Grants NNX15AD95G/NEXSS, NNX15AC89G, and NSF AST-1518332. This research has made use of data reprocessed as part of the ALICE program, which was supported by NASA through grants HST-AR-12652 (PI: R. Soummer), HST-GO-11136 (PI: D.~Golimowski), HST-GO-13855 (PI: E.~Choquet), HST-GO-13331 (PI: L.~Pueyo), and STScI Director's Discretionary Research funds, and was conducted at STScI which is operated by AURA under NASA contract NAS5-26555. The input images to ALICE processing are from the recalibrated NICMOS data products produced by the Legacy Archive project, ``A Legacy Archive PSF Library And Circumstellar Environments (LAPLACE) Investigation,'' (HST-AR-11279, PI: G.~Schneider). This  work benefited from NASA's Nexus for Exoplanet System Science (NExSS) research coordination network sponsored  by NASA's Science Mission Directorate. The authors wish to recognize and acknowledge the very significant cultural role and reverence that the summit of Mauna Kea, host to the W.~M.~Keck Observatory, has always had within the indigenous Hawaiian community. We are most fortunate to have the opportunity to conduct observations from this mountain.

\appendix
\section{The LSDV Method}\label{lsdv}
When the morphological patterns of the spiral arms do not change among different epochs, the only difference is their relative azimuthal offset. This is known as ``fixed effect'' in statistics, which has been extensively studied with the classical LSDV method. LSDV generalizes the ordinary least square (OLS) method using dummy variables. In this section, we first describe the classical LSDV method, then introduce our generalization of the method by taking into account the uncertainty from input data.

\subsection{The Classical LSDV Method}
In polar coordinates, the location of spiral arms can be represented by $(\theta_{\text{PA}}\pm\delta{\theta_{\text{PA}}}, r)$ pairs, which represents the position angle (with its uncertainty) and radial separation. For a given radial separation, its position angle can be estimated from fitting Gaussian profiles at different azimuthal directions. Assume there are $E$ epochs, each has $n$ data pairs, we can use a Taylor polynomial of degree $p$ to represent the arm morphology  \citep[e.g., ][]{grady13, benisty15, reggiani18}. The classical LSDV method finds for all the data the best fit of
\begin{equation}\label{lsdv-eq}
\theta_{\text{PA}, i} =  f(r_i) = \sum_{j=1}^p c_j r_i^j + \sum_{k=1}^{E} d_kD_k(i),
\end{equation}
where the dummy variables $D_k(i)=1$ only when the $(\theta_{\text{PA},i}, r_i)$ pair is obtained from epoch $k$, and 0 otherwise. The coefficients $d$ are then the position angles of the spiral arms when $r=0$.

Let set $\mathbb{R}^{s\times t}$ contain $s$-by-$t$ real-valued matrices, if we denote the $\theta_{\text{PA}}$'s by $\pmb\Theta\in\mathbb{R}^{nE\times1}$, with $\pmb\Theta_i=\theta_{\text{PA},i}$; the $r$'s and dummy variables by $\pmb{R}\in\mathbb{R}^{nE\times(p+E)}$, with $\pmb{R}_{i(\cdot)}=\left[r_i, r_i^2, \cdots, r_i^p, D_1(i), D_2(i), \cdots, D_E(i)\right]$; and the coefficients $\pmb{\beta}=\left[c_1,\cdots, c_p, d_1, \cdots, d_E\right]^T\in\mathbb{R}^{(p+E)\times1}$. We now write Eq.~\eqref{lsdv-eq} in a matrix OLS form:
\begin{equation}\label{lsdv-eq-matrix}
\pmb{\Theta}=\pmb{R}\pmb{\beta} +\pmb{\epsilon},
\end{equation}
where $\pmb{\epsilon}\in\mathbb{R}^{nE\times1}$ is the residual. Its cost function,
\begin{align}
C(\pmb{\Theta}, \pmb{R}; \pmb{\beta}) &= \pmb{\epsilon}^T\pmb{\epsilon}\nonumber\\
			&=\left(\pmb{\Theta}-\pmb{R}\pmb{\beta}\right)^T\left(\pmb{\Theta}-\pmb{R}\pmb{\beta}\right)\label{ssr}\\
			&= \sum_{i=1}^{nE}\left[ \theta_{\text{PA}, i} -\left( \sum_{j=1}^p c_j r_i^j + \sum_{k=1}^{E} d_kD_k(i)\right)\right]^2,\nonumber
\end{align} is minimized by
\begin{equation}\label{ols-sol}
\hat{\pmb{\beta}}=(\pmb{R}^T\pmb{R})^{-1}\pmb{R}^T\pmb{\Theta},
\end{equation}
where ${}^T$ and ${}^{-1}$ stand for matrix transpose and inverse. The standard deviations of $\hat{\pmb{\beta}}$ are calculated from the element-wise square root of the diagonal elements in the variance-covariance matrix:
\begin{equation}
\delta\hat{\pmb{\beta}}=\sqrt{\delta^2\hat{\pmb{\beta}}}=\sqrt{\text{diag}\left\{E\left[(\hat{\pmb{\beta}} -\pmb{\beta})(\hat{\pmb{\beta}} -\pmb{\beta})^T\right]\right\}} = \sqrt{\text{diag}\left\{\hat{\sigma}^2 \left(\pmb{R}^T\pmb{R} \right)^{-1}\right\}},\label{ols-sol-uncertainty}
\end{equation}
where $\hat{\sigma}^2=\frac{\left(\pmb{\Theta}-\pmb{R}\hat{\pmb{\beta}}\right)^T\left(\pmb{\Theta}-\pmb{R}\hat{\pmb{\beta}}\right)}{nE-(p+E)}$.

\subsection{The Weighted LSDV Method}
To take into account the measurement uncertainty in our study, we generalize the classical LSDV method into a weighted form. The weighted LSDV method minimizes the chi-squared statistic:
\begin{align}
\chi^2(\pmb{\Theta}, \pmb{R}; \pmb{\beta}) &= \left(\frac{\pmb{\Theta}-\pmb{R}\pmb{\beta}}{\delta\pmb{\Theta}}\right)^T\left(\frac{\pmb{\Theta}-\pmb{R}\pmb{\beta}}{\delta\pmb{\Theta}}\right)\label{chi2-cost}\\
				&=\sum_{i=1}^{nE}\left[ \frac{\theta_{\text{PA}, i}}{\delta_{\theta_{\text{PA}, i}}} -\left( \sum_{j=1}^p c_j \frac{r_i^j}{\delta_{\theta_{\text{PA}, i}}} + \sum_{k=1}^{E} d_k\frac{D_k(i)}{\delta_{\theta_{\text{PA}, i}}}\right)\right]^2\nonumber,
\end{align} 
where the division operation is element-wise; and $\delta\pmb\Theta\in\mathbb{R}^{nE\times1}$ stores the uncertainty for $\pmb\Theta$. With substitution
\begin{equation}\begin{cases}
\theta_{\text{PA}, i}' = \frac{\theta_{\text{PA}, i}}{\delta_{\theta_{\text{PA}, i}}},\\
{r'}_i^j =\frac{r_i^j}{\delta_{\theta_{\text{PA}, i}}},\\
{D}'_k(i)=\frac{D_k(i)}{\delta_{\theta_{\text{PA}, i}}},\\
\end{cases}
\end{equation}
where $'$ denotes the (element-wise) division of $\delta{\theta_{\text{PA},i}}$, we have a matrix form of \begin{align}
\chi^2(\pmb{\Theta}, \pmb{R}; \pmb{\beta}) &=\sum_{i=1}^{nE}\left[ {\theta'}_{\text{PA}, i} -\left( \sum_{j=1}^p c_j {r'}_i^j + \sum_{k=1}^{E} d_kD'_k(i)\right)\right]^2 \nonumber\\
			&= \left(\pmb\Theta'-\pmb{R}'\pmb\beta\right)^T\left(\pmb\Theta'-\pmb{R}'\pmb\beta\right)\nonumber\\
			&= C(\pmb{\Theta}', \pmb{R}'; \pmb{\beta}),
\end{align} 
as in Eq.~\eqref{ssr}, whose best-fit values and standard deviations can thus be obtained from Eqs.~\eqref{ols-sol} and \eqref{ols-sol-uncertainty}.

With two epochs of observations in our fitting, we have $E=2$ and focus only on the difference of the last two terms in $\hat{\pmb{\beta}}$, i.e., $\Delta d = \hat{d}_1-\hat{d}_2$, as well as the uncertainty $\delta(\Delta d)=\sqrt{\delta^2\hat{d}_1+\delta^2\hat{d}_2-2\text{Cov}(\hat{d}_1, \hat{d}_2)}$. In our fitting efforts, we investigated Taylor polynomials up to $p=3$ degrees to study different morphological pattern of the spiral arms, however no significant difference was observed, we therefore only report the linear results in this letter.

\end{CJK*}
\end{document}